\documentclass[aps,prd,twocolumn,amsmath,showpacs,amssymb,superscriptaddress,nofootinbib,longbibliography]{revtex4-1}
\usepackage{graphicx}
\usepackage{bm}
\usepackage{amssymb,amsmath}
\usepackage{latexsym}
\usepackage{color}
\usepackage[normalem]{ulem} 
\usepackage{dcolumn}
\usepackage[colorlinks=true,citecolor=blue,urlcolor=blue]{hyperref}

\allowdisplaybreaks

\newcommand{\Caltech}{\affiliation{Theoretical Astrophysics 350-17,
    California Institute of Technology, Pasadena, CA 91125}}

\newcommand{\CITA}{\affiliation{CITA, 60 St. George Street, Toronto, ON, M5S 3H8}}

\newcommand{\ba}{\begin{align}}
\newcommand{\ea}{\end{align}}
\newcommand{\bma}{\begin{pmatrix}}
\newcommand{\ema}{\end{pmatrix}}

\begin{document}

\title{Reply to ``On the branching of quasinormal resonances of near-extremal Kerr black holes'' by Shahar Hod}

\author{Aaron Zimmerman}
\CITA 
\author{Huan Yang}
\affiliation{Perimeter Institute for Theoretical Physics, Waterloo, Ontario N2L2Y5, Canada}
\affiliation{Institute for Quantum Computing, University of Waterloo, Waterloo, Ontario N2L3G1, Canada}
\author{Fan Zhang}
\affiliation{Gravitational Wave and Cosmology Laboratory, Department of Astronomy, Beijing Normal University, Beijing 100875, China}
\affiliation{\mbox{Department of Physics and Astronomy, West Virginia University, PO Box 6315, Morgantown, WV 26506, USA}}
\author{David A.\ Nichols}
\affiliation{Cornell Center for Astrophysics and Planetary Science (CCAPS), Cornell University, Ithaca, New York 14853, USA}
\author{Emanuele Berti}
\affiliation{Department of Physics and Astronomy, The University of Mississippi, University, MS 38677, USA}
\affiliation{CENTRA, Departamento de F\'isica, Instituto Superior T\'ecnico, Universidade de Lisboa, Avenida Rovisco Pais 1, 1049 Lisboa, Portugal}
\author{Yanbei Chen}
\Caltech
\date{\today}

\begin{abstract}
In a study of the quasinormal mode frequencies of nearly extremal black holes, we pointed out a bifurcation of the mode spectrum into modes with finite decay and modes with vanishing decay in the extremal limit. We provided analytic and semi-analytic results identifying which families of modes bifurcated, and when modes with finite decay rates exist when approaching the extremal limit. 
In a recent note~\cite{Hod:2015swa}, Hod suggests that additional modes asymptote to finite decay at extremely high spin parameter, based on past work by Detweiler. We search for these suggested modes and find no evidence of their existence. In addition, we point out an inconsistency in the derivation of the proposed modes, which further indicates that these damped modes do not exist.
\end{abstract}


\maketitle
 
\section{Introduction}
In the studies~\cite{Yang:2012pj,Yang:2013uba} we discussed the existence of two distinct families of quasinormal modes (QNMs) of nearly-extremal Kerr (NEK) black holes. In the Schwarzschild spacetime, the QNM frequencies at fixed angular quantum numbers $(l,m)$ are indexed by an overtone number $n$, and have monotonically increasing decay rates with increasing $n$. As the angular momentum of the black hole is increased, the QNM frequencies move around in the complex plane. For nearly extremal black holes some of these QNMs asymptote to finite frequencies and decay rates, while others approach the critical frequency for superradiance $\omega = m \Omega_H + O(\sqrt{\epsilon})$, where $\Omega_H$ is the angular frequency of the horizon, $\epsilon = 1 - a \ll 1$ in the regime of interest, and $a$ is the spin parameter of the black hole (here we set $G=c=M=1$). In particular, the decay rate of these modes vanishes in the extremal limit, and we called them the zero-damped modes (ZDMs). We found that whether or not some overtones remain damped in the extremal limit depends on the angular quantum numbers of the mode. Defining $\mu = m/(l+ 1/2)$, the condition in the high-frequency limit is $\mu <\mu_*$, with $\mu_* \approx 0.74$. This condition is approximately obeyed at lower frequencies as well~\cite{Yang:2012pj,Yang:2013uba}.

In his note~\cite{Hod:2015swa}, Hod suggests that even in the case $\mu \gtrsim \mu_*$, modes with finite decay exist in the limit $\epsilon \to 0$. 
His argument is based on the original analysis of Detweiler~\cite{Detweiler1980}, which in fact predicts a finite decay rate for all modes (see e.g.~\cite{Sasaki1990}), and gives a different prediction for the QNM frequencies in the NEK limit than~\cite{Hod2008a}. 
In our original studies we found no evidence for such modes. 
In this Reply to~\cite{Hod:2015swa}, we extend our analysis to the extremely small values ($\epsilon \lesssim 10^{-9}$) where Hod predicts that the QNMs should obey Detweiler's analytic prediction and asymptote to finite decay. 
As a particular example, for the case of $(s,l,m) = (2,2,2)$, Eqs.~(7) and~(8) of~\cite{Hod:2015swa} predict the QNM frequencies to be 
\begin{align}
\omega & = m \Omega_H + \varpi_n \,, 
\end{align}
with
\begin{align}
\label{eq:DetweilerPred}
\varpi_n \approx (0.162 - i \, 0.035) e^{-1.532 n} \,,
\end{align}
a value first suggested to us by S. Hod in previous communication. We reiterate that this prediction for $\omega$ has a leading-order part which is independent of $\epsilon$, although supposedly this prediction only holds at very small values of $\epsilon$.

\section{Numerical Exploration}

We have searched for this behavior in the case of $(s,l,m) = (2,2,2)$ in two ways, which we discuss below. 
Both methods rely on our implementation of Leaver's continued fraction (CF) method~\cite{Leaver1985}.
We numerically search for the zeroes of a continued fraction as we vary $\omega$, and we get accurate results for high spin parameters by actually varying the rescaled frequency $\tilde \omega = (\omega - m/2)/\sqrt{\epsilon}$.

\begin{figure}[t]
\includegraphics[width=\columnwidth]{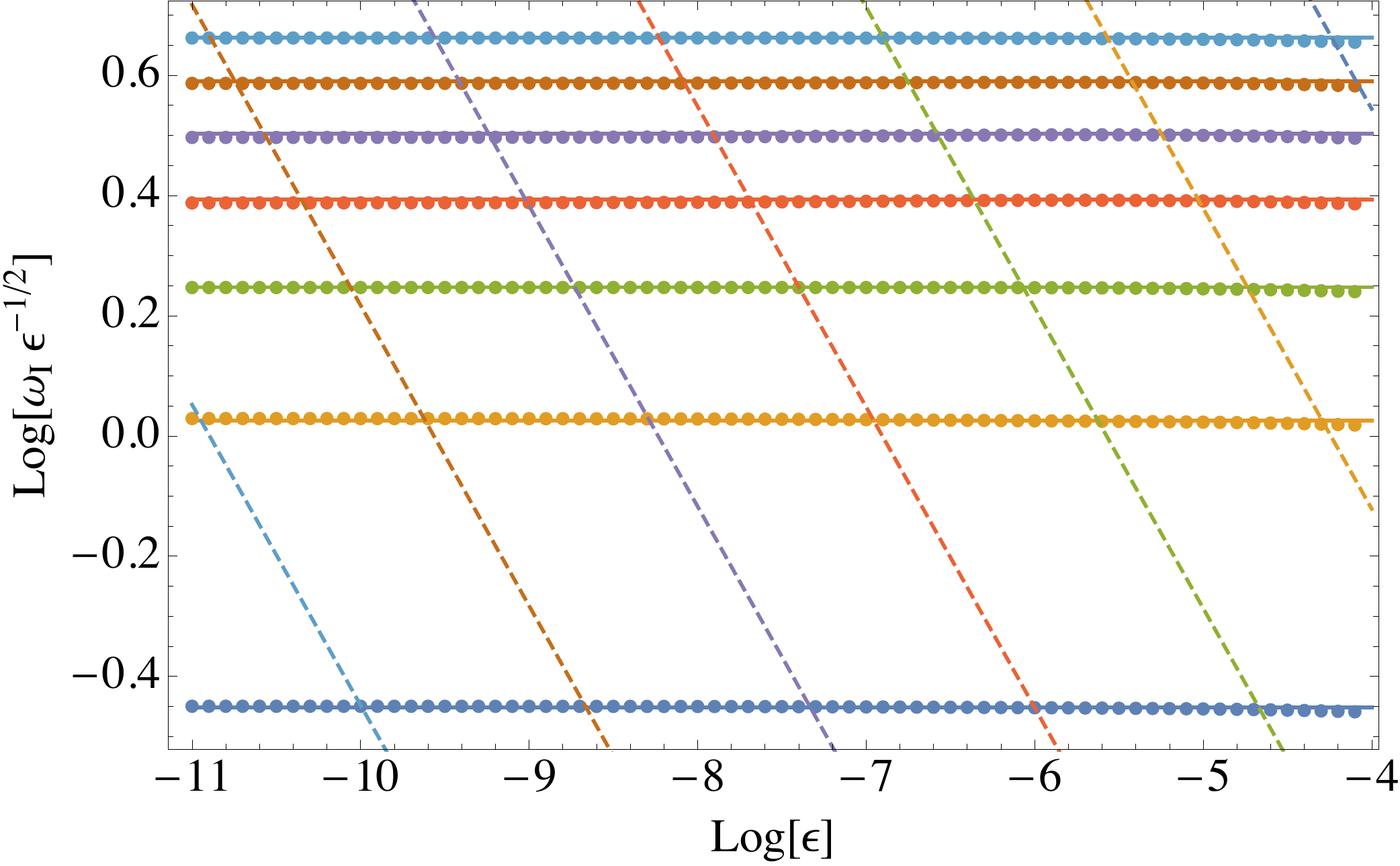}
\caption{QNM frequencies at high spins for the first seven overtones of the $(2,2,2)$ QNMs, scaled by the asymptotic behavior $\sqrt{\epsilon}$. The set of points with the lowest decay rates (blue) correspond to the $n=0$ QNMs at each $\epsilon$, and each successive set of points corresponds to $n=1$ through $n=6$. The solid lines correspond to the analytic NEK approximation discussed in ~\cite{Hod2008a,Yang:2012pj,Yang:2013uba}. We also plot the proposed decay rates from Eq.~\eqref{eq:DetweilerPred} (dashed lines), from $n=0$ (upper right) to $n = 5$ (lower left), which vary with $\epsilon$ only because of our scaling in this plot.}
\label{fig:QNMtracks}
\end{figure}

Our first method is to track the frequencies of the first seven overtones of the $(2,2,2)$ mode as we increase the spin parameter from a value well below where Hod predicts new mode behavior ($a=1-10^{-4}$) to a value well above ($a = 1 - 10^{-11}$). 
The frequencies at a given spin parameter are used iteratively to seed the frequency search at the next higher value.
Alternatively, we seed the search using our analytic approximation~\cite{Hod2008a,Yang:2012pj,Yang:2013uba} at each value of $a$ when our search jumps to another family of overtones. 
In this way, we have no difficulty identifying the QNMs and tracking them as we increase $a$ to nearly extremal values. 
None of the modes asymptotes to a fixed value of $\omega_I$ during this search, even though these QNMs pass through the values given by Eq.~\eqref{eq:DetweilerPred} in the regime of high spin parameter. In fact, all the modes are well fit by the analytic NEK predictions of~\cite{Hod2008a,Yang:2012pj,Yang:2013uba}, as illustrated in Fig.~\ref{fig:QNMtracks}. 
Note that we use the convention $\omega = \omega_R - i \omega_I$, so that $\omega_I$ is positive when the modes decay.

This search indicates that if the QNMs discussed by Hod exist, they are not the NEK limit of the QNMs that exist at lower values of $a$.
Rather, they must spontaneously appear at some large value of the spin parameter, for example by emerging from a branch cut in the complex $\omega$ plane. This behavior seems unlikely given the analytic dependence of the Teukolsky equations on $a$~\cite{PressTeukolsky1973}, but see~\cite{Hartle1974}.

\begin{figure}[t]
\includegraphics[width=1.0\columnwidth]{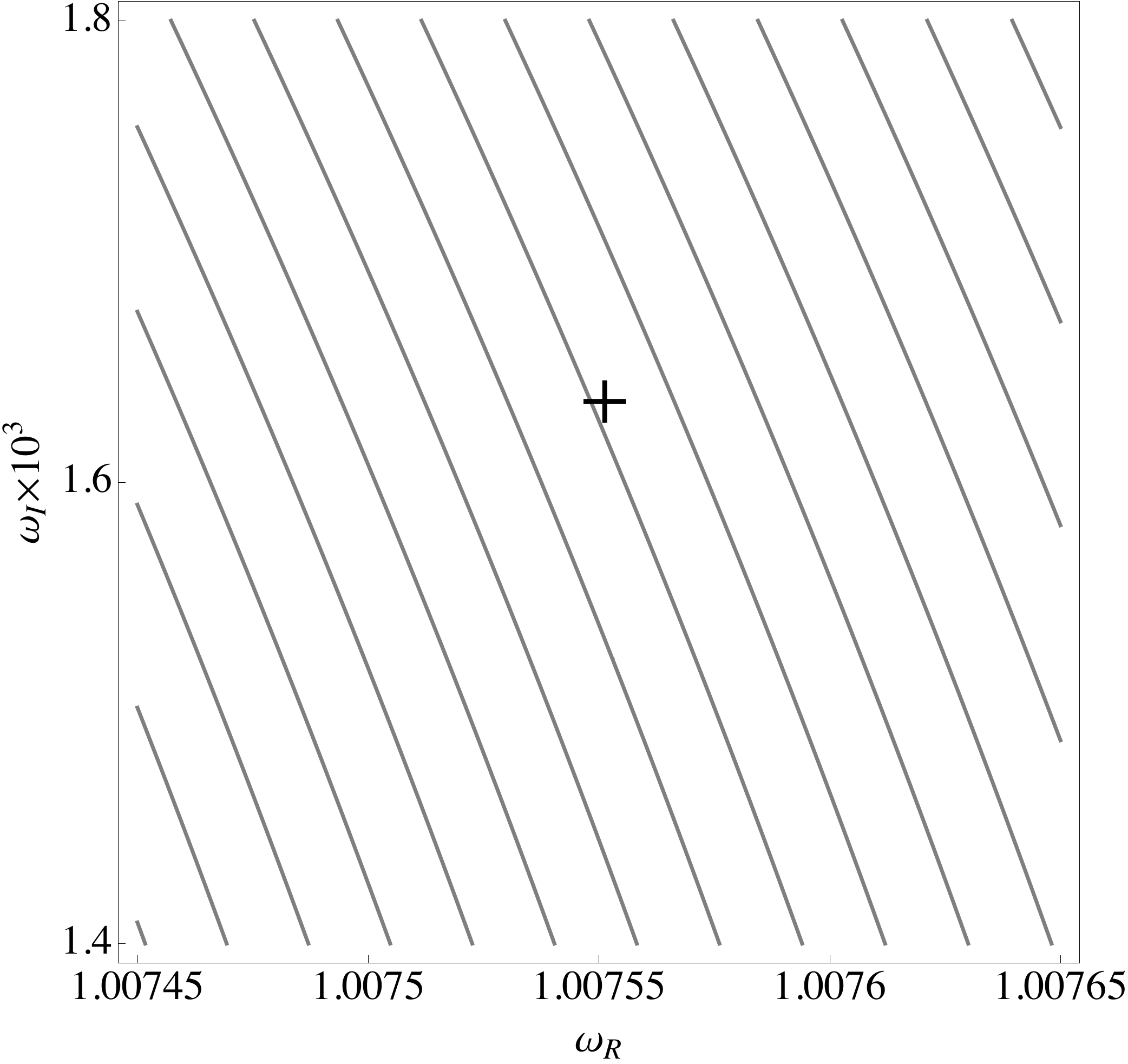}
\caption{Contours of constant magnitude of the continued fraction in the complex $\omega = \omega_R - i \omega_I$ plane for the $(2,2,2)$, $n=2$ (second overtone) QNM. The spin parameter  is $a = 1-10^{-10}$. The contours range from $10^{3.638}$ (lower left) to $10^{3.653}$ (upper right) in logarithmic steps of $0.001$. The $+$ marks the postulated DM frequency for $n=2$.}
\label{fig:CFplot}
\end{figure}

Our second method, which we found useful in our original studies, is to evaluate the CF throughout a region of the complex plane, and plot contours of constant CF value. 
The QNMs appear in regions  where the CF is approximately zero, with closed contours clustering around these points. 
In Fig.~\ref{fig:CFplot} we show the absolute value of the CF at the spin parameter $a =1-10^{-10}$ in a region near the $n=2$ QNM frequency predicted by Hod, $\omega = 1.00755 - i \, 0.00163$. 
We see no sign of a QNM. 
The CF values are $\sim 10^{3.6}$, which should be compared to the CF values of $\lesssim10^{-14}$ for the QNM frequencies of Fig.~\ref{fig:QNMtracks} at this value of $a$. 
Note that Fig.~\ref{fig:CFplot} is densely sampled, with $400\times400$ points plotted. 
In order to make such a dense sampling with a reasonable computational cost, we augment our continued fraction with a high-order and highly accurate analytic approximation for ${}_sA_{lm}$~\cite{Berti:2005gp}. We have also computed the CF in the region around the $n= 2$ prediction for $a =1-10^{-9}$, and similarly for the $n=0$ predictions at these two values of the spin parameter, with no sign of a QNM. 

\section{Analytical Argument}

The absence of the DMs discussed in Hod's Comment raises the concern that there is an error in the derivation of the frequencies in~\cite{Detweiler1980}.
A careful review of this derivation reveals a possible issue. As discussed originally in~\cite{TeukolskyPress1974}, the QNM frequencies are found in the nearly extremal case by a matching condition between two asymptotic solutions.
The first (inner) solution is found in a region where $(r - r_+)/r_+ \ll1$ using a scaled radius variable $z = - (r - r_+)/(r_+ - r_-)$. The second (outer) solution is found where $x = (r-r_+)/r_+ \gg \sqrt{\epsilon}$. These two solutions have a region of overlap where $\sqrt{\epsilon} \ll x \ll 1$.
The outgoing wave condition fixes the outer solution up to an overall normalization, and consistency with the large $|z|$ limit of the inner solution constrains the QNM\footnote{We take this opportunity to point that in the matching calculation of \cite{Yang:2013uba} the signs in front of the factors of $s$ in Eqs.~(3.9) and~(3.16) are incorrect and must be reversed; however the resulting Eq.~(3.17) is correct.}.  

\begin{figure}[t]
\includegraphics[width=1.0\columnwidth]{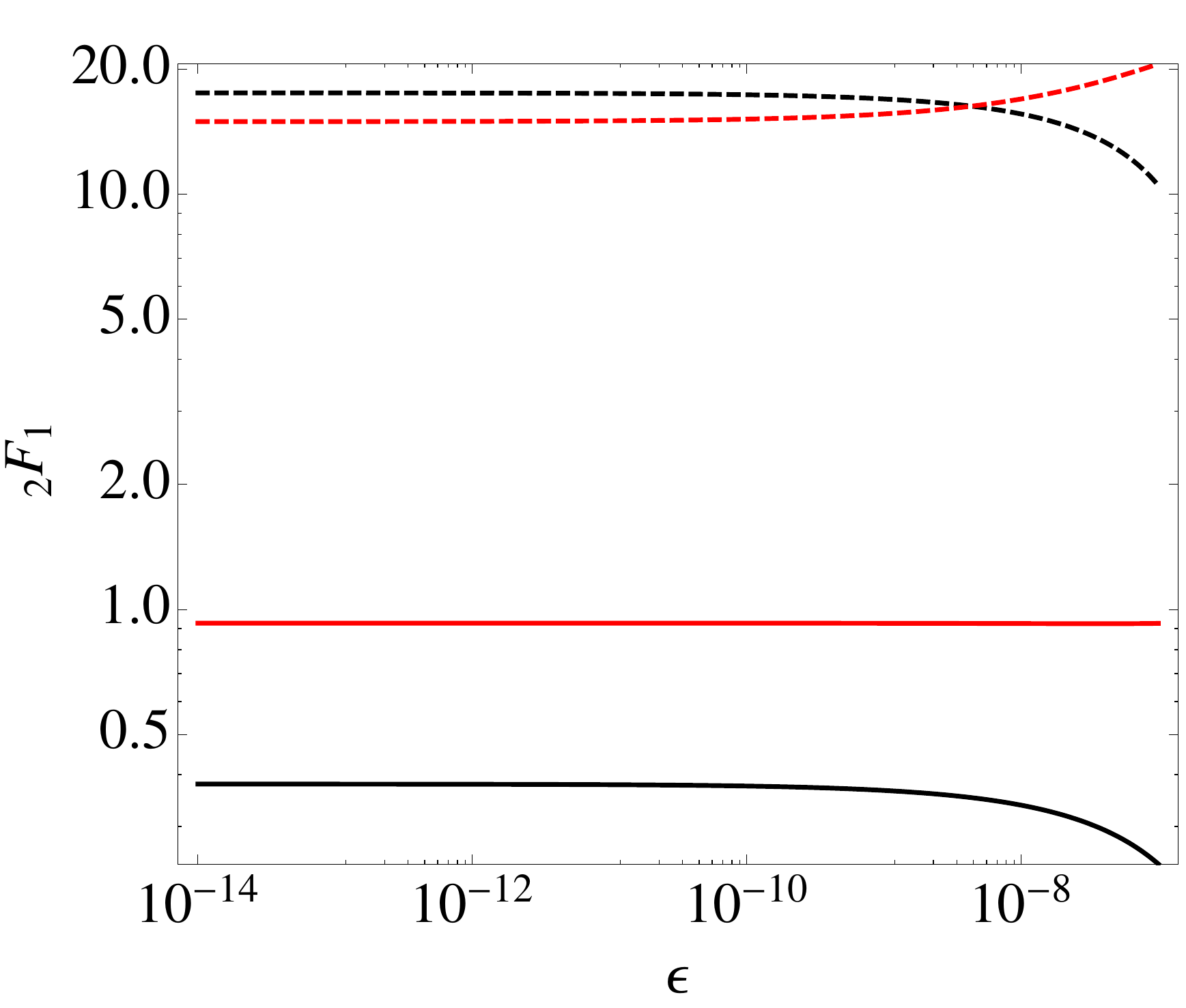}
\caption{Plot of the two hypergeometric functions in Eq.~\eqref{eq:InnerMatch} assuming the scalings in the Comment, with $l=m=n=s=2$. The solid lines represent the magnitudes of the real (black) and imaginary (red) part of ${}_2F_1\left(\alpha, \alpha - \gamma +1, \alpha - \beta, z^{-1}\right)$. The dashed lines represent the magnitudes of the real (black) and imaginary (red) part of ${}_2 F_1\left(\beta, \beta - \gamma+1, \beta - \alpha, z^{-1} \right)$. The different scaling behaviors make the matching impossible.}
\label{fig:Wavefunc}
\end{figure}

With the standard normalization that the amplitude of the QNM wavefunction is unity at the horizon, the inner solution has the form 
\begin{align}
R_{lm\omega}  =(-z)^{\kappa/2-s}(1-z)^{-i\kappa/2 - 2 i r_+ \omega - s} {}_2 F_1(\alpha, \beta, \gamma,  z)\,,
\end{align}
where $\alpha = -2 i r_+ \omega -s + 1/2 +i \delta$, $\beta = -2 i r_+ \omega -s + 1/2  -i \delta $, $\gamma = 1 - s + i \kappa$ and $\kappa = -4 r_+ (\omega - m \Omega_H)/(r_+ -r_-)$. Here $\delta$ is an analogue for the angular eigenvalue; $\delta^2 >0$ for $\mu \gtrsim \mu_*$ and $\delta^2 < 0 $ for $\mu \lesssim \mu_*$. 
In the large negative $z$ limit, the solution becomes~\cite{nist}
\begin{align}
\label{eq:InnerMatch}
{}_2 F_1(\alpha, \beta, \gamma, z) \to & \frac{\Gamma(\gamma)\Gamma(\beta - \alpha)}{\Gamma(\gamma - \alpha) \Gamma(\beta)} (-z)^{-\alpha} \notag \\ & \times {}_2F_1\left(\alpha, \alpha - \gamma +1, \alpha - \beta, z^{-1}\right) \notag \\
& + \frac{\Gamma(\gamma) \Gamma(\alpha - \beta)}{\Gamma(\gamma - \beta)\Gamma(\alpha)} (-z)^{-\beta} \notag \\ & \times {}_2 F_1\left(\beta, \beta - \gamma+1, \beta - \alpha, z^{-1} \right) \,.
\end{align}
The above formula should be compared to Eq.~(A9) in~\cite{TeukolskyPress1974}, where this matching is first discussed, and where both hypergeometric functions are taken to be $1$ in the $z \rightarrow \infty$ limit. In~\cite{Detweiler1980} and~\cite{Hod:2015swa}, the assumption is made that $\omega - m \Omega_H$ becomes finite while $r_+ -r_-$ decreases to zero, which means that $\kappa \to \infty$. However, this invalidates the matching procedure, since now in addition to $z^{-1} \rightarrow 0$, $|\gamma|$ also asymptotes to $\infty$. As a result, the functions ${}_2F_1\left(\alpha, \alpha - \gamma +1, \alpha - \beta, z^{-1}\right)$ and ${}_2 F_1\left(\beta, \beta - \gamma+1, \beta - \alpha, z^{-1} \right)$ no longer asymptote to $1$, which conflicts the matching requirement imposed in~\cite{TeukolskyPress1974,Yang:2013uba} and assumed in~\cite{Detweiler1980}. In other words, the asymptotic form of QNMs assumed in \cite{Detweiler1980,Hod:2015swa} violates the condition for their baseline equation (Eq.~(9) of \cite{Detweiler1980}) to hold in the first place.  
If one applies that baseline equation in the wrong range of validity, i.e.\ with $\omega - m \Omega_H$ remaining finite in the $a \rightarrow 1$ limit, one can find the spurious solutions in \cite{Hod:2015swa}. Figure~\ref{fig:Wavefunc} shows that when $\kappa \sim 1/\sqrt{\epsilon}$ and $z \sim r_0 /\sqrt{\epsilon}$ (with $r_0 = 0.001$) both approach large values, then the two hypergeometric functions have the wrong asymptotic limits for the matching.

\section{Conclusions}
We conclude that there is no evidence for the proposed damped modes when $\mu \gtrsim \mu_*$. The matching calculation which leads to these modes appears to be flawed, and a direct search using Leaver's method gives no evidence for these modes. 

There is still room for further understanding on how the DMs arise in the $\mu \lesssim \mu_*$ case.
These DMs cannot be treated through the matching analysis that reveals the ZDMs, since their frequencies are not near the horizon frequency $\Omega_H$ in the extremal limit.
They can be understood qualitatively through a WKB analysis~\cite{Yang:2012he}, which indicates that these modes are associated with a region outside of the near-horizon region where the ZDMs are primarily localized~\cite{Andersson2000,Yang:2012pj,Yang:2013uba}. 
The WKB analysis gives a good prediction for the lowest overtone DM in each case, and high order approximations can improve this. Presently, numerical searches are the only way to accurately compute all of the DMs in each case where $\mu \lesssim \mu_*$. 
We hope that continued interest in the problem leads to new analytic techniques and further understanding of these modes in the future. 
In particular, open questions include how many modes of a given $(s,l,m)$ asymptote to finite decay in the extremal limit, and at what spin the spectrum bifurcates into two branches.

\acknowledgements
We thank S. Hod for discussing an earlier version of his work with us. 
A.Z. was supported by the Beatrice and Vincent Tremaine Postdoctoral Fellowship at the Canadian Institute for Theoretical Astrophysics.
H.Y.  acknowledges support from the Perimeter Institute of Theoretical Physics and the Institute for Quantum Computing. Research at Perimeter Institute is supported by the government of Canada and by the Province of Ontario though Ministry of Research and Innovation.
F.Z. acknowledges support from the National Natural Science Foundation of China (Grants 11443008 and 11503003), the Fundamental Research Funds for the Central Universities (Grant No.~2015KJJCB06), and a Returned Overseas Chinese Scholars Foundation grant. 
D.N. is supported by  NSF grants PHY-1404105 and PHY-1068541.
E.B. was supported by NSF CAREER Grant No.~PHY-1055103 and by FCT contract IF/00797/2014/CP1214/CT0012 under the IF2014 Programme.
Y.C. is supported by NSF grant PHY-1404569 and CAREER Grant PHY-0956189.

\bibliography{}

\begin{thebibliography}{14}%
\makeatletter
\providecommand \@ifxundefined [1]{%
 \@ifx{#1\undefined}
}%
\providecommand \@ifnum [1]{%
 \ifnum #1\expandafter \@firstoftwo
 \else \expandafter \@secondoftwo
 \fi
}%
\providecommand \@ifx [1]{%
 \ifx #1\expandafter \@firstoftwo
 \else \expandafter \@secondoftwo
 \fi
}%
\providecommand \natexlab [1]{#1}%
\providecommand \enquote  [1]{``#1''}%
\providecommand \bibnamefont  [1]{#1}%
\providecommand \bibfnamefont [1]{#1}%
\providecommand \citenamefont [1]{#1}%
\providecommand \href@noop [0]{\@secondoftwo}%
\providecommand \href [0]{\begingroup \@sanitize@url \@href}%
\providecommand \@href[1]{\@@startlink{#1}\@@href}%
\providecommand \@@href[1]{\endgroup#1\@@endlink}%
\providecommand \@sanitize@url [0]{\catcode `\\12\catcode `\$12\catcode
  `\&12\catcode `\#12\catcode `\^12\catcode `\_12\catcode `\%12\relax}%
\providecommand \@@startlink[1]{}%
\providecommand \@@endlink[0]{}%
\providecommand \url  [0]{\begingroup\@sanitize@url \@url }%
\providecommand \@url [1]{\endgroup\@href {#1}{\urlprefix }}%
\providecommand \urlprefix  [0]{URL }%
\providecommand \Eprint [0]{\href }%
\providecommand \doibase [0]{http://dx.doi.org/}%
\providecommand \selectlanguage [0]{\@gobble}%
\providecommand \bibinfo  [0]{\@secondoftwo}%
\providecommand \bibfield  [0]{\@secondoftwo}%
\providecommand \translation [1]{[#1]}%
\providecommand \BibitemOpen [0]{}%
\providecommand \bibitemStop [0]{}%
\providecommand \bibitemNoStop [0]{.\EOS\space}%
\providecommand \EOS [0]{\spacefactor3000\relax}%
\providecommand \BibitemShut  [1]{\csname bibitem#1\endcsname}%
\let\auto@bib@innerbib\@empty
\bibitem [{\citenamefont {Hod}(2015)}]{Hod:2015swa}%
  \BibitemOpen
  \bibfield  {author} {\bibinfo {author} {\bibfnamefont {Shahar}\ \bibnamefont
  {Hod}},\ }\bibfield  {title} {\enquote {\bibinfo {title} {{On the branching
  of the quasinormal resonances of near-extremal Kerr black holes}},}\
  }\href@noop {} {\  (\bibinfo {year} {2015})},\ \Eprint
  {http://arxiv.org/abs/1510.05604} {arXiv:1510.05604 [gr-qc]} \BibitemShut
  {NoStop}%
\bibitem [{\citenamefont {Yang}\ \emph
  {et~al.}(2013{\natexlab{a}})\citenamefont {Yang}, \citenamefont {Zhang},
  \citenamefont {Zimmerman}, \citenamefont {Nichols}, \citenamefont {Berti}
  \emph {et~al.}}]{Yang:2012pj}%
  \BibitemOpen
  \bibfield  {author} {\bibinfo {author} {\bibfnamefont {Huan}\ \bibnamefont
  {Yang}}, \bibinfo {author} {\bibfnamefont {Fan}\ \bibnamefont {Zhang}},
  \bibinfo {author} {\bibfnamefont {Aaron}\ \bibnamefont {Zimmerman}}, \bibinfo
  {author} {\bibfnamefont {David~A.}\ \bibnamefont {Nichols}}, \bibinfo
  {author} {\bibfnamefont {Emanuele}\ \bibnamefont {Berti}},  \emph {et~al.},\
  }\bibfield  {title} {\enquote {\bibinfo {title} {{Branching of quasinormal
  modes for nearly extremal Kerr black holes}},}\ }\href {\doibase
  10.1103/PhysRevD.87.041502} {\bibfield  {journal} {\bibinfo  {journal}
  {Phys.Rev.}\ }\textbf {\bibinfo {volume} {D87}},\ \bibinfo {pages} {041502}
  (\bibinfo {year} {2013}{\natexlab{a}})},\ \Eprint
  {http://arxiv.org/abs/1212.3271} {arXiv:1212.3271 [gr-qc]} \BibitemShut
  {NoStop}%
\bibitem [{\citenamefont {Yang}\ \emph
  {et~al.}(2013{\natexlab{b}})\citenamefont {Yang}, \citenamefont {Zimmerman},
  \citenamefont {Zengino{\u g}lu}, \citenamefont {Zhang}, \citenamefont {Berti}
  \emph {et~al.}}]{Yang:2013uba}%
  \BibitemOpen
  \bibfield  {author} {\bibinfo {author} {\bibfnamefont {Huan}\ \bibnamefont
  {Yang}}, \bibinfo {author} {\bibfnamefont {Aaron}\ \bibnamefont {Zimmerman}},
  \bibinfo {author} {\bibfnamefont {Anıl}\ \bibnamefont {Zengino{\u g}lu}},
  \bibinfo {author} {\bibfnamefont {Fan}\ \bibnamefont {Zhang}}, \bibinfo
  {author} {\bibfnamefont {Emanuele}\ \bibnamefont {Berti}},  \emph {et~al.},\
  }\bibfield  {title} {\enquote {\bibinfo {title} {{Quasinormal modes of nearly
  extremal Kerr spacetimes: spectrum bifurcation and power-law ringdown}},}\
  }\href {\doibase 10.1103/PhysRevD.88.044047} {\bibfield  {journal} {\bibinfo
  {journal} {Phys.Rev.}\ }\textbf {\bibinfo {volume} {D88}},\ \bibinfo {pages}
  {044047} (\bibinfo {year} {2013}{\natexlab{b}})},\ \Eprint
  {http://arxiv.org/abs/1307.8086} {arXiv:1307.8086 [gr-qc]} \BibitemShut
  {NoStop}%
\bibitem [{\citenamefont {{Detweiler}}(1980)}]{Detweiler1980}%
  \BibitemOpen
  \bibfield  {author} {\bibinfo {author} {\bibfnamefont {S.}~\bibnamefont
  {{Detweiler}}},\ }\bibfield  {title} {\enquote {\bibinfo {title} {{Black
  holes and gravitational waves. III - The resonant frequencies of rotating
  holes}},}\ }\href {\doibase 10.1086/158109} {\bibfield  {journal} {\bibinfo
  {journal} {\apj}\ }\textbf {\bibinfo {volume} {239}},\ \bibinfo {pages}
  {292--295} (\bibinfo {year} {1980})}\BibitemShut {NoStop}%
\bibitem [{\citenamefont {{Sasaki}}\ and\ \citenamefont
  {{Nakamura}}(1990)}]{Sasaki1990}%
  \BibitemOpen
  \bibfield  {author} {\bibinfo {author} {\bibfnamefont {M.}~\bibnamefont
  {{Sasaki}}}\ and\ \bibinfo {author} {\bibfnamefont {T.}~\bibnamefont
  {{Nakamura}}},\ }\bibfield  {title} {\enquote {\bibinfo {title}
  {{Gravitational radiation from an extreme Kerr black hole}},}\ }\href
  {\doibase 10.1007/BF00756835} {\bibfield  {journal} {\bibinfo  {journal}
  {General Relativity and Gravitation}\ }\textbf {\bibinfo {volume} {22}},\
  \bibinfo {pages} {1351--1366} (\bibinfo {year} {1990})}\BibitemShut {NoStop}%
\bibitem [{\citenamefont {{Hod}}(2008)}]{Hod2008a}%
  \BibitemOpen
  \bibfield  {author} {\bibinfo {author} {\bibfnamefont {S.}~\bibnamefont
  {{Hod}}},\ }\bibfield  {title} {\enquote {\bibinfo {title} {{Slow relaxation
  of rapidly rotating black holes}},}\ }\href {\doibase
  10.1103/PhysRevD.78.084035} {\bibfield  {journal} {\bibinfo  {journal}
  {\prd}\ }\textbf {\bibinfo {volume} {78}},\ \bibinfo {eid} {084035} (\bibinfo
  {year} {2008})},\ \Eprint {http://arxiv.org/abs/0811.3806} {arXiv:0811.3806
  [gr-qc]} \BibitemShut {NoStop}%
\bibitem [{\citenamefont {Leaver}(1985)}]{Leaver1985}%
  \BibitemOpen
  \bibfield  {author} {\bibinfo {author} {\bibfnamefont {E.W.}\ \bibnamefont
  {Leaver}},\ }\bibfield  {title} {\enquote {\bibinfo {title} {{An Analytic
  representation for the quasi normal modes of Kerr black holes}},}\
  }\href@noop {} {\bibfield  {journal} {\bibinfo  {journal} {Proc. Roy. Soc.
  Lond. A}\ }\textbf {\bibinfo {volume} {402}},\ \bibinfo {pages} {285--298}
  (\bibinfo {year} {1985})}\BibitemShut {NoStop}%
\bibitem [{\citenamefont {{Press}}\ and\ \citenamefont
  {{Teukolsky}}(1973)}]{PressTeukolsky1973}%
  \BibitemOpen
  \bibfield  {author} {\bibinfo {author} {\bibfnamefont {W.~H.}\ \bibnamefont
  {{Press}}}\ and\ \bibinfo {author} {\bibfnamefont {S.~A.}\ \bibnamefont
  {{Teukolsky}}},\ }\bibfield  {title} {\enquote {\bibinfo {title}
  {{Perturbations of a Rotating Black Hole. II. Dynamical Stability of the Kerr
  Metric}},}\ }\href {\doibase 10.1086/152445} {\bibfield  {journal} {\bibinfo
  {journal} {\apj}\ }\textbf {\bibinfo {volume} {185}},\ \bibinfo {pages}
  {649--674} (\bibinfo {year} {1973})}\BibitemShut {NoStop}%
\bibitem [{\citenamefont {{Hartle}}\ and\ \citenamefont
  {{Wilkins}}(1974)}]{Hartle1974}%
  \BibitemOpen
  \bibfield  {author} {\bibinfo {author} {\bibfnamefont {J.~B.}\ \bibnamefont
  {{Hartle}}}\ and\ \bibinfo {author} {\bibfnamefont {D.~C.}\ \bibnamefont
  {{Wilkins}}},\ }\bibfield  {title} {\enquote {\bibinfo {title} {{Analytic
  properties of the Teukolsky equation}},}\ }\href {\doibase
  10.1007/BF01651548} {\bibfield  {journal} {\bibinfo  {journal}
  {Communications in Mathematical Physics}\ }\textbf {\bibinfo {volume} {38}},\
  \bibinfo {pages} {47--63} (\bibinfo {year} {1974})}\BibitemShut {NoStop}%
\bibitem [{\citenamefont {Berti}\ \emph {et~al.}(2006)\citenamefont {Berti},
  \citenamefont {Cardoso},\ and\ \citenamefont {Casals}}]{Berti:2005gp}%
  \BibitemOpen
  \bibfield  {author} {\bibinfo {author} {\bibfnamefont {Emanuele}\
  \bibnamefont {Berti}}, \bibinfo {author} {\bibfnamefont {Vitor}\ \bibnamefont
  {Cardoso}}, \ and\ \bibinfo {author} {\bibfnamefont {Marc}\ \bibnamefont
  {Casals}},\ }\bibfield  {title} {\enquote {\bibinfo {title} {{Eigenvalues and
  eigenfunctions of spin-weighted spheroidal harmonics in four and higher
  dimensions}},}\ }\href {\doibase 10.1103/PhysRevD.73.109902,
  10.1103/PhysRevD.73.024013} {\bibfield  {journal} {\bibinfo  {journal} {Phys.
  Rev.}\ }\textbf {\bibinfo {volume} {D73}},\ \bibinfo {pages} {024013}
  (\bibinfo {year} {2006})},\ \bibinfo {note} {[Erratum: Phys.
  Rev.D73,109902(2006)]},\ \Eprint {http://arxiv.org/abs/gr-qc/0511111}
  {arXiv:gr-qc/0511111 [gr-qc]} \BibitemShut {NoStop}%
\bibitem [{\citenamefont {{Teukolsky}}\ and\ \citenamefont
  {{Press}}(1974)}]{TeukolskyPress1974}%
  \BibitemOpen
  \bibfield  {author} {\bibinfo {author} {\bibfnamefont {S.~A.}\ \bibnamefont
  {{Teukolsky}}}\ and\ \bibinfo {author} {\bibfnamefont {W.~H.}\ \bibnamefont
  {{Press}}},\ }\bibfield  {title} {\enquote {\bibinfo {title} {{Perturbations
  of a rotating black hole. III - Interaction of the hole with gravitational
  and electromagnetic radiation}},}\ }\href {\doibase 10.1086/153180}
  {\bibfield  {journal} {\bibinfo  {journal} {\apj}\ }\textbf {\bibinfo
  {volume} {193}},\ \bibinfo {pages} {443--461} (\bibinfo {year}
  {1974})}\BibitemShut {NoStop}%
\bibitem [{\citenamefont {Olver}\ \emph {et~al.}(2010)\citenamefont {Olver},
  \citenamefont {Lozier}, \citenamefont {Boisvert},\ and\ \citenamefont
  {Clark}}]{nist}%
  \BibitemOpen
  \bibfield  {author} {\bibinfo {author} {\bibfnamefont {Frank~W.}\
  \bibnamefont {Olver}}, \bibinfo {author} {\bibfnamefont {Daniel~W.}\
  \bibnamefont {Lozier}}, \bibinfo {author} {\bibfnamefont {Ronald~F.}\
  \bibnamefont {Boisvert}}, \ and\ \bibinfo {author} {\bibfnamefont
  {Charles~W.}\ \bibnamefont {Clark}},\ }\href@noop {} {\emph {\bibinfo {title}
  {NIST Handbook of Mathematical Functions}}},\ \bibinfo {edition} {1st}\ ed.\
  (\bibinfo  {publisher} {Cambridge University Press},\ \bibinfo {address} {New
  York, NY, USA},\ \bibinfo {year} {2010})\BibitemShut {NoStop}%
\bibitem [{\citenamefont {Yang}\ \emph {et~al.}(2012)\citenamefont {Yang},
  \citenamefont {Nichols}, \citenamefont {Zhang}, \citenamefont {Zimmerman},
  \citenamefont {Zhang} \emph {et~al.}}]{Yang:2012he}%
  \BibitemOpen
  \bibfield  {author} {\bibinfo {author} {\bibfnamefont {Huan}\ \bibnamefont
  {Yang}}, \bibinfo {author} {\bibfnamefont {David~A.}\ \bibnamefont
  {Nichols}}, \bibinfo {author} {\bibfnamefont {Fan}\ \bibnamefont {Zhang}},
  \bibinfo {author} {\bibfnamefont {Aaron}\ \bibnamefont {Zimmerman}}, \bibinfo
  {author} {\bibfnamefont {Zhongyang}\ \bibnamefont {Zhang}},  \emph {et~al.},\
  }\bibfield  {title} {\enquote {\bibinfo {title} {{Quasinormal-mode spectrum
  of Kerr black holes and its geometric interpretation}},}\ }\href {\doibase
  10.1103/PhysRevD.86.104006} {\bibfield  {journal} {\bibinfo  {journal}
  {Phys.Rev.}\ }\textbf {\bibinfo {volume} {D86}},\ \bibinfo {pages} {104006}
  (\bibinfo {year} {2012})},\ \Eprint {http://arxiv.org/abs/1207.4253}
  {arXiv:1207.4253 [gr-qc]} \BibitemShut {NoStop}%
\bibitem [{\citenamefont {{Andersson}}\ and\ \citenamefont
  {{Glampedakis}}(2000)}]{Andersson2000}%
  \BibitemOpen
  \bibfield  {author} {\bibinfo {author} {\bibfnamefont {N.}~\bibnamefont
  {{Andersson}}}\ and\ \bibinfo {author} {\bibfnamefont {K.}~\bibnamefont
  {{Glampedakis}}},\ }\bibfield  {title} {\enquote {\bibinfo {title}
  {{Superradiance Resonance Cavity Outside Rapidly Rotating Black Holes}},}\
  }\href {\doibase 10.1103/PhysRevLett.84.4537} {\bibfield  {journal} {\bibinfo
   {journal} {Physical Review Letters}\ }\textbf {\bibinfo {volume} {84}},\
  \bibinfo {pages} {4537--4540} (\bibinfo {year} {2000})},\ \Eprint
  {http://arxiv.org/abs/arXiv:gr-qc/9909050} {arXiv:gr-qc/9909050} \BibitemShut
  {NoStop}%
\end{thebibliography}%

\end{document}